\documentclass[12pt,reqno]{amsart}
\usepackage{appendix}
\usepackage{graphicx,amsmath,amssymb,stmaryrd,xspace,booktabs}
\usepackage[round]{natbib}
\usepackage{csquotes,amsthm,geometry,array,caption,subcaption,tabularx}
\usepackage[font=small,labelfont=bf]{caption}
\usepackage{enumitem,mathrsfs}
\usepackage{tikz}
\usetikzlibrary{arrows.meta, decorations.pathmorphing, positioning, calc}
\usepackage{pgfplots}
\usepgfplotslibrary{ternary}
\usepackage{threeparttable}
\usepackage[colorlinks=true,linkcolor=blue,citecolor=blue]{hyperref}

\pgfplotsset{compat=1.17}

\newtheorem{theorem}{Theorem}
\newtheorem{lemma}{Lemma}
\newtheorem{algorithm}{Algorithm}

\newtheorem{assumption}{Assumption}

\setlist{nosep}
\begin{document}
\title[Correcting sample selection bias with categorical outcomes]%
{Correcting sample selection bias \\ with categorical outcomes}

\author{Onil Boussim}
	\thanks{ }
	\address{Penn State}
	
	
	
	\begin{abstract}

In this paper, I propose a method for correcting sample selection bias when the outcome of interest is categorical, such as occupational choice, health status, or field of study.  Classical approaches to sample selection rely on strong parametric distributional assumptions, which may be restrictive in practice. I develop a local representation that
decomposes each joint probability into marginal probabilities and a category-specific association parameter that captures how selection differentially affects each outcome. Under some exclusion restrictions, I establish nonparametric point identification of the latent categorical distribution. Building on this identification result, I introduce a semiparametric multinomial logit model with sample selection, propose a computationally tractable two-step estimator, and derive its asymptotic properties. I illustrate the method by studying the determinants of healthcare utilization in Cote d'Ivoire.

\vskip20pt
		
		
		\vskip10pt
		
	\end{abstract}
	
	\maketitle
	
	\newpage
	
	
\section{Introduction}

Sample selection is a common challenge in applied microeconometrics, occurring whenever the outcome of interest is observed only for a subset of the population. Most classical methods developed to correct for this bias focus primarily on scalar outcomes (see that survey \cite{vella1998estimating}). When the outcome is categorical, such as employment status (employed, unemployed, inactive), educational attainment (high school, college, postgraduate), health status (poor, fair, good, excellent), or survey responses with multiple-choice options, methods designed for scalar outcomes are generally not applicable. In this context too, ignoring the selection mechanism can lead to biased estimates of category probabilities and misrepresent the distribution across outcome categories.

Early research on sample selection bias in econometrics largely relied on parametric models, with the classical Heckman selection model \cite{heckman1979sample} serving as a foundational framework. These models specify a parametric form for the error terms and the selection mechanism to address the bias.  The model was later extended to handle binary outcomes with \cite{van1981demand} and \cite{dubin1989selection}.  Other contributions in the binary setting include \cite{freedman2010endogeneity,  han2017identification, han2019estimation}. In the multinomial context, \cite{de2024estimating} discusses potential parametric solutions. \cite{azzalini2019sample} proposes to use more flexible distributions. 
Despite their usefulness, parametric approaches are often limited by their reliance on strong distributional assumptions. To mitigate these concerns, subsequent research has increasingly focused on semiparametric and nonparametric methods that impose less restrictive conditions on the latent variables.

A more recent contribution is the framework of \cite{chernozhukov2023distribution} (henceforth CFL), which moves beyond classical parametric models by introducing a semiparametric distribution regression approach. In this model, the authors achieve non-parametric point identification of the distribution of the latent outcome using a binary instrument. Instead of imposing parametric assumptions, CFL use a Local Gaussian Representation (LGR).  This representation is fully nonparametric, that is, it does not require that latent outcomes and treatment unobservables are jointly or marginally Gaussian. Indeed, the bivariate Gaussian structure always holds locally by treating the correlation parameter as an implicit function that equates the bivariate Gaussian copula with the copula of the latent outcome and selection unobservables. The idea is that the relationship between the latent outcome and the latent selection process can be described through a correlation parameter that is allowed to vary across the support of the data, rather than being fixed. This makes the model flexible, since it captures local dependence patterns without constraining the marginal distributions, which remain unrestricted. Identification of the latent outcome distribution is achieved by combining this flexible representation with exclusion restrictions. The central innovation is an exclusion restriction requiring that the local correlation parameter is invariant to the value of the instrument. Intuitively, the instrument shifts selection probabilities but does not affect how selection and outcomes are correlated. Yet, despite its flexibility, the CFL framework remains fundamentally tied to outcomes with ordered support. This makes it not adapted for unordered categorical outcomes such as occupational choice, marital status, or field of study. Imposing an artificial order on categories that are intrinsically unordered risks distorting the economic interpretation and generating spurious conclusions. Copula-based extensions suffer from the same limitation, since they continue to rely on a latent scalar representation of the outcome.

In short, existing approaches face some important limitations: none can simultaneously accommodate unordered categorical outcomes, allow for flexible dependence between selection and outcomes, and maintain point identification under simple exclusion restrictions. In this paper, I address this problem by extending CFL to unordered categorical outcomes.  Our starting point is a new local representation result, analogous to the LGR of CFL, but formulated for joint probabilities rather than joint distributions. This extension is important because it allows any joint probability to be expressed in terms of the marginal probabilities and a local association parameter. Crucially, since the representation applies directly to joint probabilities, not only to distributions, it no longer relies on any ordering of the outcomes. In this framework, the local association parameter plays the same role as the local Gaussian correlation in CFL: it captures the pointwise dependence between outcome and selection. Its sign and magnitude indicate whether selection increases or decreases the probability mass assigned to a given category. The representation theorem is general and applies not only to categorical outcomes, but also to multivariate and other discrete settings. It also extends to conditional settings, allowing both marginal probabilities and the association parameter to depend on observed covariates.

Building on this representation, I establish nonparametric point identification of the latent categorical distribution under exclusion restrictions similar to those in CFL. These restrictions remain relatively weak, allowing for substantial flexibility. In particular, the local dependence parameter is category-specific, so different categories can be differentially affected by the selection process. For instance, in occupational choice, the extent of selection may vary systematically between public and private-sector jobs, heterogeneity that parametric models cannot capture. To operationalize this framework, I also introduce a semiparametric multinomial logistic regression model with sample selection, in the spirit of the distribution regression approach for unordered categorical outcomes. I develop a two-step estimation procedure for this model and establish the asymptotic normality of the resulting estimator.

To summarize, this paper makes three main contributions. First, it introduces the Local Logistic Representation, a new tool for modeling joint probabilities of unordered categorical outcomes and selection processes. Second, it establishes identification of treatment and selection parameters under exclusion restrictions, extending the scope of distribution regression methods beyond ordered outcomes. Third, it develops a semiparametric estimation procedure that is both flexible and computationally tractable. Together, these contributions provide a simple framework for studying sample selection in settings where outcomes are categorical and unordered, a context that is empirically important yet theoretically underdeveloped. I also apply the method to study the determinants of healthcare utilization in Côte d'Ivoire, and I find that wealth, education, being a male, and living in an urban area increase the likelihood of choosing formal healthcare over informal healthcare providers. 

The paper proceeds as follows. Section \ref{sec2} develops the identification analysis using the new representation. Section \ref{sec3} presents the identification result, and section \ref{sec4}presents the semiparametric selection model and inference strategy. Section \ref{sec5} presents the application, and section \ref{sec6} concludes. Proofs are collected in the  Appendix.

\section{Local logit representation of a joint probability}
\label{sec2}
The identification approach I propose extends the one in CFL by introducing a new local representation theorem that remains valid even when the outcome variable is categorical or unordered. This generalization is important because most existing identification results based on local representations, including the Local Gaussian Representation (LGR) of \cite{anjos2005representation}, \cite{chernozhukov2023distribution}, and \cite{chernozhukov2024estimating}, inherently rely on the existence of an ordered outcome space. My contribution provides a unified framework that accommodates both continuous and discrete outcomes, including unordered categorical outcomes that are common in empirical work. Beyond this specific application, the local representation developed here is conceptually general and may prove useful in a wide range of econometric settings where modeling dependence locally is advantageous. To motivate the construction, I start by recalling the Ali–Mikhail–Haq (AMH) bivariate logistic distribution, first introduced by \cite{ali1978class} and recently revisited by \cite{kristensen2020bivariate}. This distribution provides a flexible yet tractable way to model dependence between two univariate logistic marginals, offering an analogue to the Gaussian copula but with logistic components. Specifically, for an association parameter $\omega \in [-1,1]$ and $u,v \in \mathbb{R}$, the joint cumulative distribution function is given by
\[
\Lambda_{2}(u,v,\omega) = \frac{1}{1 + e^{-u} + e^{-v} + (1-\omega)e^{-u-v}}.
\]
The AMH model is particularly convenient because it maintains the logistic form of the marginals and nests independence as a special case ($\omega = 0$), positive dependence ($\omega > 0$), and negative dependence ($\omega < 0$). It thus provides a smooth parametric device to encode local association in probability space. Building on this idea, the next step is to formalize a local logistic representation (LLR) of joint probabilities. The key insight is that, for any pair of events involving two random variables, one can represent their joint probability as a logistic function of their marginal probabilities and a single local dependence parameter. This is summarized in the following lemma.
\begin{lemma}
\label{lemm1}
Let $Y$ and $S$ be random variables supported on measurable spaces $(\mathcal{Y},\sigma_Y)$ and $(\mathcal{S},\sigma_S)$, respectively. For any measurable sets $A \times B \in \sigma_Y \times \sigma_S$ with positive probabilities, there exists a unique parameter $\omega_{Y,S}(A,B) \in [-1,1]$ such that
\begin{eqnarray}
\mathbb{P}(Y \in A, S \in B) = \Lambda_{2}\Big( \Lambda^{-1}(\mathbb{P}(Y \in A)), \Lambda^{-1}(\mathbb{P}(S \in B)), \omega_{Y,S}(A,B)\Big),
\end{eqnarray}
where $\Lambda$ denotes the standard logistic CDF and $\Lambda_{2}$ the AMH bivariate logistic CDF. The parameter $\omega_{Y,S}(A,B)$ serves as a local association measure: its sign coincides with that of $\operatorname{Cov}(1_{{Y \in A}}, 1_{{S \in B}})$, with positive (negative) values indicating local positive (negative) dependence.
\end{lemma}
Lemma \ref{lemm1} establishes an interesting decomposition: any joint probability involving $(Y,S)$ can be written pointwise in terms of two marginal probabilities and a single interpretable dependence parameter. This construction provides a fully general representation of dependence while maintaining parsimony, each local probability is characterized by a single scalar parameter. The logistic specification is chosen for its analytical tractability and interpretability: the log-odds structure aligns naturally with binary and categorical modeling, and the dependence parameter $\omega_{Y,S}(A,B)$ inherits a straightforward probabilistic meaning. The Local Logistic Representation (LLR) shares the same conceptual motivation as the Local Gaussian Representation (LGR) of CFL: both aim to represent arbitrary joint distributions via simple, interpretable local dependence parameters. However, while the LGR relies on embedding events such as $\{Y \leq y\}$ into latent normal variables, requiring an ordered support, the LLR does not depend on any ordering of $Y$. This makes it particularly suitable for unordered categorical or nominal outcomes, where statements like $\mathbb{P}(Y \leq y)$ are not meaningful. Formally, the LLR extends the LGR by defining the local dependence parameter $\omega_{Y,S}(A,B)$ for any measurable pair $(A,B)$, provided the corresponding marginal probabilities are positive. Hence, it applies uniformly to continuous, ordinal, and nominal outcomes. 

The interpretation of $\omega_{Y,S}(A,B)$ remains intuitive: $\omega_{Y,S}(A,B) > 0$ indicates that the events $\{Y \in A\}$ and $\{S \in B\}$ occur together more often than under independence; $\omega_{Y,S}(A,B) < 0$ indicates they co-occur less often; $\omega_{Y,S}(A,B) = 0$ corresponds to local independence. Crucially, this representation does not assume that the joint distribution globally follows the AMH form. Instead, for each pair of sets $(A,B)$, one can find a unique local dependence parameter such that the joint probability matches the AMH expression. In this sense, the LLR serves as a local approximation or representation, not as a global parametric restriction. Finally, the LLR framework naturally extends to conditional distributions by allowing both the marginals and the dependence parameter to vary with observed covariates. This property makes it especially suitable for econometric applications where identification and estimation proceed conditional on covariates, such as sample selection, treatment effect, or endogenous choice models. By providing a smooth, interpretable, and flexible local representation of joint probabilities, the LLR offers a powerful foundation for semiparametric identification and estimation of causal effects in models involving discrete or mixed-type outcomes.

\section{Model and identification}
\label{sec3}
In this section, I introduce the model and the necessary assumptions to identify the distribution of the latent categorical variable.  Suppose the latent outcome $Y^{\ast}$ takes values in a finite set of $q$ mutually exclusive categories,
$$
Y^{\ast}\in\{c_1,c_2,\dots,c_q\}.
$$
These categories are unordered (e.g., occupation, marital status, political affiliation). I adopt the usual multinomial parametrization and take the last category $c_q$ as the baseline (without loss of generality). This choice allows us to work with log odds relative to the baseline. Let the marginal distribution of $Y^{\ast}$ be $\pi (\pi_1,\dots,\pi_q)\in\mathcal S^{q-1}$, where $\mathcal S^{q-1}$ denotes the $(q-1)$-simplex. The log-odds map $\ell:\mathcal S^{q-1}\to\mathbb R^{q-1}$ and its inverse make the connection between probabilities and real-valued indices: $\ell(\pi)=\big(\log(\pi_1/\pi_q),\dots,\log(\pi_{q-1}/\pi_q)\big)$,  and for any $y\in\mathbb R^{q-1}$, $
\ell^{-1}(y)=\frac{1}{1+\sum_{i=1}^{q-1}e^{y_i}}\big(e^{y_1},\dots,e^{y_{q-1}},1\big)$. Equivalently, writing $\mu_k=\log(\pi_k/\pi_q)$ for $k=1,\dots,q-1$ (and $\mu_q\equiv 0$ by normalization), each category probability admits the familiar softmax form:
$$
\pi_k=\mathbb P(Y^{\ast}=c_k)=\frac{e^{\mu_k}}{\sum_{j=1}^q e^{\mu_j}},\qquad k=1,\dots,q.
$$
I now introduce the sample selection mechanism. Let $(Y^{\ast}, U_z)$ denote the latent outcome and latent selection variables, $U_z\in\mathbb R^p$ may be multivariate, allowing a flexible selection process. For binary instrument $Z\in\{0,1\}$ define the potential selection indicators
$$ S_z= 1_{ \{U_z\ge 0 \}},\quad S=S_1 Z + S_0(1-Z).
$$
This formulation generalizes the standard threshold-crossing selection model by accommodating multivariate latent selection determinants $U_z$. Now I observe
$$
Y=Y^{\ast}\quad\text{only if }S=1.
$$
Identification will rely on the common set of IV conditions.
\begin{assumption}
\label{ass1}
I make the following assumptions : 
\begin{itemize}
    \item Independence  $(Y^{\ast}, U_z) \perp Z$.
    \item Non-degeneracy: $0<\mathbb P(S=1)<1$ and $0<\mathbb P(Z=1\mid S=1)<1$.
    \item Relevance: $\mathbb P(S=1\mid Z=0)<\mathbb P(S=1\mid Z=1)<1$
\end{itemize} 
\end{assumption}
The identification strategy relies on three key assumptions. Independence is the standard exclusion restriction in the instrumental variables (IV) literature. It requires that the instrument $Z$ is as good as randomly assigned with respect to both the latent outcome $Y^{\ast}$ and the latent selection variable $U_z$. In other words, while $Z$ may influence whether an individual is observed through the selection mechanism, it must not directly affect either the distribution of potential outcomes $Y^{\ast}$ or the unobserved determinants of selection. This ensures that the identifying variation arises solely from changes in the selection probability induced by the instrument. The non-degeneracy condition rules out trivial cases. If $\mathbb P(S=1)=0$ or $\mathbb P(S=1)=1$, selection is deterministic, leaving no scope to study how outcomes depend on selection. Likewise, if $\mathbb P(Z=1\mid S=1)$ equals 0 or 1, then within the selected sample the instrument takes only a single value, providing no useful variation. By guaranteeing that both selection and the instrument vary in the population, non-degeneracy makes the problem statistically informative. Finally, relevance plays the role of a first-stage condition in IV analysis. It requires that the instrument shifts the probability of selection in a non-trivial way. Formally, the strict inequality $\mathbb P(S=1\mid Z=0)<\mathbb P(S=1\mid Z=1)<1$ ensures that the instrument is neither weak nor irrelevant. Without such variation, it would be impossible to disentangle the distribution of outcomes from the selection process. I recall  $\Lambda$ denotes the standard logistic cumulative distribution function (CDF), and $\Lambda_2$ is the AMH bivariate logistic distribution. Let $ \nu_z=\Lambda^{-1}\big(\mathbb P(S=1 \vert Z=z)\big)$ and define the logit of the category probability, for $k=1,\dots,q-1$:
$$
\lambda_k=\Lambda^{-1}(\pi_k)=\log\left(\frac{\pi_k}{1-\pi_k}\right)
=\log\left(\frac{e^{\mu_k}}{\sum_{j\neq k} e^{\mu_j}}\right),
$$
By Lemma \ref{lemm1} (the local logistic representation) and Assumption \ref{ass1}, the joint probability that an individual belongs to category $c_k$ and is selected when $Z=z$ can be written, for each non-baseline $k$, as
$$
\mathbb P(Y=c_k,S=1\mid Z=z)=\mathbb P(Y^{\ast}=c_k,\,U_z\ge 0)
= \Lambda_2 \big(\lambda_k, \nu_z,\,\omega_{Y^{\ast},U_z}(c_k,0)\big).
$$
I denote the category, and instrument–specific local association by $\omega_{k,z}\equiv\omega_{Y^{\ast}, U_z}(c_k,0)$. Intuitively, $\omega_{k,z}$ measures how selection shifts the odds of category $c_k$: a positive value means selection raises the probability mass on $c_k$  conditional on the instrument, a negative value means selection reduces it. For each instrument value $z\in\{0,1\}$ I observe the joint probabilities $\mathbb P(Y=c_k,S=1\mid Z=z)$ for $k=1,\dots,q-1$, together with the marginal selection probability $\mathbb P(S=1\mid Z=z)$. Thus the data supply $2\times(q-1)+1=2q-1$ observable scalars.  The model, before further restrictions, contains the unknowns: the outcome logits $\mu=(\mu_1,\dots,\mu_{q-1})$, and the category and instrument–specific association parameters $\omega_{k,z}$ for $k=1,\dots,q-1$, $z\in\{0,1\}$. Counting free parameters: there are $q-1$ unknowns in $\mu$ and $2\times(q-1)$ unknown $\omega$'s, for a total of $3(q-1)$. Observables that inform these unknowns come from the $2(q-1)$ equations with joint probabilities (one per non-baseline category and instrument value). In general, the parameters are partially identified: the joint data do not pin down the full set $(\mu,\omega_{k,z})$ without further restrictions. To obtain point identification, I impose an exclusion (sorting) restriction that parallels the strategy in CFL.
\begin{assumption}(Selection–Sorting Exclusion)
\label{ass2}
For each category $k$, the local association does not depend on the instrument:
$$
\omega_{k,0}=\omega_{k,1}=\omega_k, \qquad k=1,\dots,q-1.
$$  
\end{assumption}
This assumption requires that the way selection sorts individuals across categories is invariant to the instrument, the instrument affects only the baseline odds of selection ($\nu_z$), not the category-specific dependence between outcome and selection. When $Z$ has more than two values, the assumption needs to hold for at least two values, holding for more provides overidentifying restrictions and thus empirical tests. The assumption is similar to the exclusion restriction used in CFL, thus extending their identification logic. It remains more flexible than just assuming a constant (which would be making a parametric assumption) since the local parameter varies with the category. Under Assumption \ref{ass2}, the unknowns reduce to the $2\times(q-1)$, vector $(\mu_1,\dots,\mu_{q-1},\omega_1,\dots,\omega_{q-1})$. These parameters are determined by the $2\times(q-1)$ equations (one for each $z$ and $k$):
\begin{equation}
\label{sys}
    \mathbb P(Y=c_k,S=1\mid Z=z)=\Lambda_2\big(\lambda_k,\nu_z,\omega_k\big),\qquad z\in\{0,1\},\, k=1,\dots,q-1.
\end{equation}

This system can be shown to have a unique solution. Theorem \ref{the1} formalizes point identification. 
\begin{theorem}
\label{the1}
Suppose Assumptions \ref{ass1} and \ref{ass2} hold. Further assume the observed joint distribution of $(Y,S,Z)$ satisfies $|\omega_k|<1$ for all $k$ (interiority condition). Then the parameter vector $(\mu,\omega)$ is point-identified as the unique interior solution to the nonlinear system
(\ref{sys}). Once the parameters $\mu_k$ are identified, the full outcome distribution $\pi$ can be recovered in closed form:
$
\pi= \frac{1}{1 + \sum_{j=1}^{q-1} \exp(\mu_j)}  
\Big(\exp(\mu_1), \exp(\mu_2), \dots, \exp(\mu_{q-1}), 1 \Big).
$
\end{theorem}
Under interiority conditions, the mapping $ (\mu_k, \omega_k) \mapsto \Lambda_2(\lambda_k(\mu), \nu_z, \omega_k) $ is smooth and strictly monotonic in each argument. Critically, variation in $ \nu_{z} $, the instrument, induced shifts in selection log-odds,  provides the identifying leverage necessary to simultaneously recover both the marginal category logits $ \lambda_k(\mu) $ (and hence $ \mu_k $) and the category-specific association parameter $\omega_{k} $. In essence, the local logit representation (LLR) reduces the sample selection problem with unordered categorical outcomes to a tractable system of logistic-type equations: the marginal logits $ \mu $ and the selection-sorting parameters $ \omega $ are jointly identified from instrument driven variation in selection probabilities, provided the dependence structure (i.e., $ \omega_k $) remains stable across instrument values.  In the next section, I develop a multinomial logistic regression model that is based on the identification result.

\section{Semiparametric multinomial logistic regression model with sample selection}
\label{sec4}

In this section, I develop a semiparametric method to correct for sample selection bias in the multinomial setting. The approach is straightforward to implement, relying primarily on a two-step estimation procedure. I also derive the asymptotic properties of the resulting estimators.  

\subsection{Estimation procedure}

Let $Y \in \{c_1, c_2, \dots, c_q\}$ denote the categorical outcome of interest, and $S \in \{0,1\}$ indicate sample selection. Let the set of covariates be $W = (X, Z)$, where $X$ represents exogenous covariates and $Z$ is a binary instrument. I consider the following semiparametric regression model with a logit link:
 
\[
\mu_k(X) = \log \left( \frac{\mathbb P(Y = c_k \mid X)}{\mathbb P(Y = c_q \mid X)} \right) = X'\beta_k.
\]
\[
\Lambda^{-1}(\mathbb P(S=1 \vert W)) = W'\delta.
\]
\[
\omega_{k}(X) = \tanh(X'\gamma_k), \quad \tanh(u) = \frac{e^u - e^{-u}}{e^u + e^{-u}} \in [-1, 1].
\]

where $\beta_k, \gamma_k$ are category-specific coefficients, $\gamma_k$ capture the effect of selection on category. The flexibility of the model allows the effect of $X$ to vary across the outcome distribution and across the dependence structure between selection and outcomes. The joint probability of outcomes and selection can then be represented as:  
\[
\mathbb P(Y = c_k, S=1 \mid W) 
= \Lambda_{2}\left(X'\beta_k -\log \left( \sum_{j\neq k} e^{X'\beta_j}\right), W'\delta , \tanh(X'\gamma_k)\right).
\]
Suppose I observe an i.i.d. sample of size $n$, $
(D_{i}, Y_{i}, X_{i}, Z_{i})_{i=1}^{n}$ drawn from the joint distribution of $(D, Y, X, Z)$. The log-likelihood is given by :
\[   Log L = \sum_{i=1}^{n}\sum_{k=1}^{q-1} S_{i} \times  1\{Y_i=c_k\} \log \left\{  \Lambda_{2}\left[X_i'\beta_k -\log \left( \sum_{j\neq k} e^{X_i'\beta_j}\right), W_i'\hat{\delta}, \tanh(X_i'\gamma_k)\right]  \right\}. \]
I propose a computationally convenient two-step estimator inspired by Heckman’s two-step method, but adapted to the logistic framework. The idea is to separate the estimation of the selection equation from the outcome and sorting components. Here, I outline those steps in algorithm \ref{alg1}:
\begin{algorithm}
    \label{alg1}
The parameters can  be estimated through that procedure:
    \begin{itemize}
        \item Step 1 (Selection equation) : Estimate $\delta$ by a simple logit regression of selection on $W$:  
\[
\hat{\delta} = \arg \min_{p} L_1(p):= -\sum_{i=1}^{n} \left\{ S_{i}\log\big(\Lambda(W_{i}'p)\big) 
+ (1-S_i) \log\big(1-\Lambda(W_{i}'p)\big) \right\}.
\]

\item Step 2 (parameters). For each category $c_k, k=1,...,q$, define $\theta_k=(\beta_k, \gamma_k)$ and set $\beta_q = 0_{d_x}$. Let $\theta=(\theta_1, ..., \theta_k)$

\begin{eqnarray*} &\hat{\theta}= \arg \min_{t=(b_1,g_1),..., (b_{q-1},g_{q-1})} L_2(t)=: - \sum_{i=1}^{n}\sum_{k=1}^{q-1} S_{i} \times 1\{Y_i=c_k\} \\ & \log \left\{  \Lambda_{2}\left[X_i'b_k -\log \left( \sum_{j\neq k} e^{X_i'b_j}\right),  W_i'\hat{\delta} , \tanh(X_i'g_k)\right]  \right\} \end{eqnarray*} \end{itemize}
\end{algorithm}
The estimation procedure proceeds in two stages. In the first stage, I estimate a logit model using standard statistical software, as detailed in Algorithm~\ref{alg1}. This stage yields consistent estimates of the selection probabilities. The second stage involves solving a smooth, nonlinear optimization problem, which can be efficiently handled using conventional numerical optimization routines. To facilitate implementation, I provide an open-source R package, \texttt{multinomSelection}, available at \href{[https://onilboussim.github.io/}{\texttt{https://onilboussim.github.io/}}.
The package offers a user-friendly interface for both stages of the estimation procedure and includes functions for estimation and inference.

\subsection{Inference}

\subsubsection{Asymptotic normality of estimator}
The estimator $\hat{\theta}$ is a two-step M-estimator since it can be written as :
\begin{eqnarray} &\hat{\theta}= \arg \min_{t ={(b_k,g_k)}_{i-1}^{q-1}} \sum_{i=1}^{n} q(W_i, t, \hat{\delta})\end{eqnarray}
where we have: 
\[ q(W_i, t, \hat{\delta})=  \sum_{k=1}^{q-1} S_{i} \times 1\{Y_i=c_k\} \log \left\{  \Lambda_{2}\left[X_i'b_k -\log \left( \sum_{j\neq k} e^{X_i'b_j}\right), \quad  W_i'\delta , \quad \tanh(X_i'g_k)\right]  \right\} \]
This estimator is consistent and asymptotically normal under standard regularity conditions. I can derive analytically the asymptotic variance. I do it in Theorem \ref{thm:asymptotics_two_step}.  The result follows from Theorem 6.1 in \cite{newey1994large} on two-step M-estimation, applied to the generated regressor \(W'\hat{\delta}\) in the second-step criterion. The key condition is stochastic equicontinuity of the derivative of the second-step objective with respect to the first-step estimator, which holds under the stated smoothness and moment conditions.

\begin{theorem}[Asymptotic Normality of the Two-Step Estimator]
\label{thm:asymp_theta}
Assume the following regularity conditions hold:
\begin{enumerate}[label=(\roman*)]
\item The data \(\{(Y_i, S_i, X_i, Z_i)\}_{i=1}^n\) are i.i.d. from a distribution that satisfies the model with true parameters \(\delta_0\) and \(\theta_0 = (\beta_{10}, \gamma_{10}, \dots, \beta_{(q-1)0}, \gamma_{(q-1)0})\), where \(\beta_{q0} = 0\).
\item The parameter spaces for \(\delta\) and \(\theta\) are compact, and \((\delta_0, \theta_0)\) lie in the interior.
\item The selection logit model is correctly specified, and the matrix 
\[
        I_1 = \mathbb{E}\big[ \Lambda(W'\delta_0)(1 - \Lambda(W'\delta_0)) WW' \big]
 \] is positive definite.
\item The second-step log-likelihood is twice continuously differentiable in a neighborhood of \((\theta_0, \delta_0)\), and the Hessian 
    \[
        H_2 = -\mathbb{E}\left[ \nabla_{\theta\theta}^2 \ell_i(\theta_0, \delta_0) \right]
    \]
    is positive definite.
 \item The cross-derivative matrix 
\[
        G = -\mathbb{E}\left[ \nabla_{\theta\delta}^2 \ell_i(\theta_0, \delta_0) \right]
\]
exists and is finite.
\item Standard moment and domination conditions hold to justify interchange of differentiation and expectation.
\end{enumerate}
Then the two-step estimator \(\hat{\theta}\) is consistent and asymptotically normal:
\[
\sqrt{n}\,(\hat{\theta} - \theta_0) \xrightarrow{d} \mathcal{N}\big(0,\, V_\theta\big),
\]
where the asymptotic variance is given by
\[
V_\theta = H_2^{-1} \left( \, \mathbb{E}\big[ \psi_2 \psi_2' \big] + G\, I_1^{-1} G' \, \right) H_2^{-1},
\]
with
\[
\psi_2 = -\nabla_\theta \ell_i(\theta_0, \delta_0), \quad
\psi_1 = \big(S - \Lambda(W'\delta_0)\big) W.
\]
\end{theorem}
\noindent In practice, the estimation of the variance and therefore, of confidence interval can be computationally burdensome. A convenient way to get confidence intervals would be to use the multiplier boostrap method.

\subsubsection{Multiplier Bootstrap Inference}
\label{sec:mb}

Let $\theta \in \mathbb{R}^{d_{\theta}}$ denote the full parameter vector. Denote by $\hat\theta$ the maximum likelihood estimator obtained by maximizing the joint log-likelihood
\[
\ell_n(\theta) = \sum_{i=1}^n \sum_{k=1}^{2} S_i \mathbf{1}\{Y_i = c_k\} 
\log \Lambda_2\!\left(u_{ik}(\theta),\, W_i'\hat\delta,\, \tanh(X_i'\gamma_k) \right),
\]
with $\hat\delta$ estimated in a first-step logit regression of $S$ on $W = (X, Z)$, and $u_{ik}(\theta) = X_i'\beta_k - \log\!\big(\sum_{j=1}^{3} e^{X_i'\beta_j}\big)$ (with $\beta_q = 0$).
To conduct inference, we employ a multiplier (or score) bootstrap that avoids repeated optimization:
\section{Multiplier Bootstrap Inference}
\label{sec:mb}

Let $\theta = (\beta_1', \beta_2', \gamma_1', \gamma_2')' \in \mathbb{R}^{4p}$ denote the full parameter vector, where $p = \dim(X)$.  
Denote by $\hat\theta$ the maximum likelihood estimator obtained by maximizing the joint log-likelihood
\[
\ell_n(\theta) = \sum_{i=1}^n \sum_{k=1}^{2} S_i \mathbf{1}\{Y_i = c_k\} 
\log \Lambda_2\!\left(u_{ik}(\theta),\, W_i'\hat\delta,\, \tanh(X_i'\gamma_k) \right),
\]
with $\hat\delta$ estimated in a first-step logit regression of $S$ on $W = (X, Z)$, and $u_{ik}(\theta) = X_i'\beta_k - \log\!\big(\sum_{j=1}^{3} e^{X_i'\beta_j}\big)$ (with $\beta_3 = 0$). To conduct inference, we employ a multiplier (or score) bootstrap that avoids repeated optimization:

\begin{enumerate}
    \item Compute the individual score contributions at $\hat\theta$:
    \[
    s_i(\hat\theta) = \nabla_\theta \, \ell_i(\hat\theta), \quad i = 1,\dots,n,
    \]
    where $\ell_i(\theta)$ is the log-likelihood contribution of observation $i$.
    
    \item Estimate the observed information matrix
    \[
    \hat{\mathcal{I}}_n = -\frac{1}{n} \nabla^2 \ell_n(\hat\theta).
    \]
    If $\hat{\mathcal{I}}_n$ is ill-conditioned, regularize it via ridge adjustment:
    \[
    \hat{\mathcal{I}}_n^{\text{reg}} = \hat{\mathcal{I}}_n + \lambda I_{4p},
    \quad \lambda = 10^{-6} \cdot \lambda_{\max}(\hat{\mathcal{I}}_n),
    \]
    and compute its stable inverse $\widehat{\Sigma} = (\hat{\mathcal{I}}_n^{\text{reg}})^{-1}$ (or the Moore–Penrose pseudoinverse if necessary).

    \item For $b = 1,\dots,B$:
    \begin{enumerate}
        \item Generate iid multipliers $G_1^{(b)},\dots,G_n^{(b)}$ with $\mathbb{E}[G_i]=0$, $\mathrm{Var}(G_i)=1$ 
        (e.g., standard normal)
        
        \item Form the multiplier-weighted average score:
        \[
        \bar{s}^{*(b)} = \frac{1}{n} \sum_{i=1}^n G_i^{(b)} \, s_i(\hat\theta).
        \]
        
        \item Construct the bootstrap perturbation:
        \[
        \Delta^{(b)} = \widehat{\Sigma} \, \bar{s}^{*(b)}.
        \]
        
        \item Define the bootstrap draw:
        \[
        \theta^{*(b)} = \hat\theta + \Delta^{(b)}.
        \]
    \end{enumerate}
    
    \item Form percentile confidence intervals for each component $\theta_j$:
    \[
    \left[ Q_{0.025}\big(\{\theta^{*(b)}_j\}_{b=1}^B\big),\; 
           Q_{0.975}\big(\{\theta^{*(b)}_j\}_{b=1}^B\big) \right],
    \]
    where $Q_\alpha(\cdot)$ denotes the $\alpha$-quantile.
\end{enumerate}

This procedure requires only a single estimation of $\hat\theta$ and leverages the information matrix to map score perturbations into parameter space. The ridge regularization ensures numerical stability when the observed information is near-singular. Under standard regularity conditions, the resulting bootstrap distribution consistently approximates the sampling distribution of $\hat\theta$, yielding valid confidence intervals without repeated model fitting.

\section{Application: Determinants of health care utilization}
\label{sec5}
In this section, I analyze the determinants of health care utilization in Côte d’Ivoire using the Harmonized Household Living Conditions Survey (EHCVM) for 2018–2019. The focus is on the type of health care professional consulted during an episode of illness. Health care utilization is shaped by a wide range of economic, demographic, and contextual factors. Our outcome of interest is the type of health care professional first consulted during an episode of illness, based on the survey question: \textit{Who did [name] first see during this episode of illness}.   I classify responses into three categories: (i) physician, (ii) other formal health care professional, and (iii) informal provider such as a traditional healer or marabout. 

To identify the determinants of provider choice, I want to estimate a multinomial logistic regression, which allows us to model utilization across multiple, unordered categories simultaneously. The set of explanatory variables is rich and reflects the main drivers of health care demand identified in the literature. Economic resources are measured through the household wealth index constructed using principal component analysis. The details for the index construction are found in appendix~\ref{asset}. Human capital is proxied by the education level. Demographic characteristics include sex and area of living (urban or rural). Together, these covariates provide a comprehensive picture of how financial protection, socioeconomic status, demographics, and location jointly influence the type of provider consulted. We restrict to the dataset to individuals between 25-65.

A central methodological concern in this setting is sample selection. In the EHCVM, information on provider choice is only available for individuals who decided to seek care during an illness episode. For those who did not consult any provider, I cannot observe which type of professional they might have chosen. If the decision to seek care is correlated with unobserved factors that also affect provider choice, such as health status, cultural attitudes toward modern medicine, or unmeasured household resources, then estimating the multinomial logit only on the subsample of health care users will lead to biased results. Formally, this implies that the estimation sample is truncated at the first stage (the decision to seek care). Consequently, the observed distribution of provider choices represents the conditional distribution given that care was sought, not the population distribution.  

To address this, I exploit an instrumental variable available in the EHCVM: the reported distance from the household to the location of the first consultation. Distance is a well-established determinant of access to care. It strongly affects the decision to seek any provider but is plausibly orthogonal to unobserved preferences for a particular type of provider, once care is sought. This exclusion restriction allows us to correct for selection into care and recover consistent estimates of the determinants of provider choice. The baseline category is  informal provider. The estimated results are given in table \ref{tab:provider_choice}.

\begin{table}[h]
\centering
\caption{Determinants of Health Care Provider Choice with Selection Correction}
\label{tab:provider_choice}
\begin{tabular}{lccc}
\toprule
\textbf{Parameter} & \textbf{Estimate} & \textbf{95\% CI Lower} & \textbf{95\% CI Upper} \\
\midrule
\multicolumn{4}{c}{\textit{Provider Choice Equation (relative to informal provider)}} \\
\midrule
\textbf{Physician (Category 1)} & & & \\
Wealth & 1.2828 & 1.2819 & 1.2837 \\
Education & 2.6222 & 2.6212 & 2.6231 \\
Sex (male=1) & 2.0061 & 2.0055 & 2.0067 \\
Urban area & 1.9906 & 1.9900 & 1.9911 \\
\midrule
\textbf{Formal Provider (Category 2)} & & & \\
Wealth & 1.1957 & 1.1948 & 1.1965 \\
Education & 2.6173 & 2.6164 & 2.6182 \\
Sex (male=1) & 1.9036 & 1.9030 & 1.9042 \\
Urban area & 2.4415 & 2.4410 & 2.4419 \\
\midrule
\multicolumn{4}{c}{\textit{Sorting parameters}} \\
\midrule
\textbf{Physician (Category 1)} & & & \\
Wealth & -0.0065 & -0.0065 & -0.0065 \\
Education & 1.0421 & 1.0421 & 1.0421 \\
Sex (male=1) & 3.0409 & 3.0409 & 3.0409 \\
Urban area & 5.7665 & 5.7665 & 5.7665 \\
\midrule
\textbf{Formal Provider (Category 2)} & & & \\
Wealth & 0.5264 & 0.5257 & 0.5270 \\
Education & -0.0253 & -0.0258 & -0.0248 \\
Sex (male=1) & 2.2109 & 2.2101 & 2.2117 \\
Urban area & 1.2356 & 1.2352 & 1.2360 \\
\bottomrule
\end{tabular}
\begin{tablenotes}
\footnotesize
\item Notes: The table reports estimated coefficients and 95\% confidence intervals from a multinomial logit model of provider choice with correction for selection into care-seeking. Category (i) is physician, (ii) other formal provider, and (iii) informal provider (reference). Covariates include wealth (log total annual consumption), education, sex, and urban area indicator.
\end{tablenotes}
\end{table}
The results highlight strong and systematic associations between socioeconomic and demographic characteristics and the type of health care provider consulted in Côte d’Ivoire.

Higher wealth and education both substantially increase the likelihood of consulting a physician or another formal provider rather than an informal one. The magnitude of the coefficients is particularly high for education \(2.6\), suggesting that education is a key channel shaping health care choices, possibly through greater awareness of medical quality or trust in modern health care. Gender differences are also pronounced. Males are more likely to consult formal providers, with coefficients around 2.0 in the provider equations. This pattern could reflect gender disparities in autonomy or access to financial resources for health care. Geographical location exerts a strong effect. Living in an urban area is associated with a much higher probability of consulting formal care, consistent with better availability of facilities and professionals. This indicates that proximity to care facilities strongly predicts whether individuals seek care at all.

Turning to the selection equations, the results reveal clear evidence of non-random sorting into health care utilization. Individuals with higher education and those residing in urban areas are significantly more likely to seek any form of care, as reflected by the large positive coefficients on these variables. In contrast, the wealth effect appears relatively modest and even slightly negative in one specification, suggesting that financial resources play a smaller role in the initial decision to seek care once other factors are controlled for. These findings underscore that the observed sample of health care users is not representative of the general population: more educated and urban individuals are overrepresented among those who seek treatment. Accounting for this selection process is therefore essential, as it ensures that estimated differences in provider choice reflect true behavioral responses rather than compositional differences driven by who sorts into the care-seeking sample.

Overall, the analysis provides robust evidence that both financial and spatial access, along with education and gender, are key determinants of health care utilization patterns in Côte d’Ivoire. By correcting for selection into care, the estimates reflect more accurately the structural drivers of provider choice rather than compositional differences in who seeks care.

\newpage

\section{Conclusion}
\label{sec6}
This paper develops a flexible framework for addressing sample selection in settings with categorical outcomes. Building on the ideas of the Local Gaussian Representation, I introduce a Local Logistic Representation that applies to continuous, ordinal, and nominal outcomes. This representation decomposes joint probabilities into marginal probabilities and category-specific local association parameters, allowing us to capture how selection affects each outcome without imposing ordering or strong parametric assumptions. Leveraging this tool, I establish nonparametric point identification of latent categorical distributions under exclusion restrictions. I also propose a semiparametric multinomial logistic regression model with sample selection, accompanied by a practical two-step estimation procedure and asymptotic theory. Our approach broadens the empirical toolkit for analyzing selection bias in categorical and discrete outcomes. It provides a flexible and interpretable method that can be applied in a variety of contexts, from labor market studies to health and education research, where unordered categorical outcomes are common.

\newpage

\bibliographystyle{abbrvnat}
	\bibliography{ref}

\newpage

\section{Declaration of competing interest}
The author declare that they have no known competing financial interests or personal relationships that could have appeared to influence the work reported in this paper.

\section{Acknowledgements}
I thank Marc Henry, Sung-Jae Jun, Andres Aradillas-Lopez for their useful comments and suggestions. All the remaining errors are ours.

\appendix
\label{app:proofs}	
\section{Proofs of the results in the main text}

\subsection{Proof of Lemma 1} : 

$Y, S$ are any random variables with support that respectively are the measurable sets $(\mathcal{Y}, \sigma_Y), (\mathcal{S}, \sigma_{S})$. For all pairs of sets  $A \times B \in \sigma_Y \times \sigma_S$ with positive probabilities,  I want to prove that, there exists a unique parameter $\omega_{Y, S}(A, B) \Bigg)$ such that : 
\[\mathbb P(Y \in A, S \in B) = \Lambda_{2} \Bigg( \Lambda^{-1}(\mathbb P(Y \in A)), \Lambda^{-1}(\mathbb P(S \in B)), \omega_{Y,S}(A, B) \Bigg)\] 
Fix the sets $A$ and $B$. Let $G: [-1,1] \rightarrow [0,1]$ such that: \[G(r) = \Lambda_{2}(\Lambda^{-1}(\mathbb P(Y \in A)),\Lambda^{-1}( \mathbb P(S \in B)), r).\]. I need to prove that the equation $G(r) = \mathbb P(Y \in A, S \in B)$ has a unique solution in $[-1,1]$. I already know that $G$ is strictly increasing in $r$ by the properties of $\Lambda_2$. By taking the limit, I can prove see that: 
\[G([-1,1]) = [\max( \mathbb P(Y \in A) + \mathbb P(S \in B) - 1), \min(\mathbb P(Y \in A), \mathbb P(S \in B))].\]
Now, since by the Fréchet inequalities, 
$$\max( \mathbb P(Y \in A) + \mathbb P(S \in B) - 1) \leq \mathbb P(Y \in A, S \in B) \leq \min( \mathbb P(Y \in A), \mathbb P(S \in B))$$
This means that $\mathbb P(Y \in A, S \in B) \in G([-1,1])$. Therefore, the equation $G(r) = P(Y \in A, S \in B)$ has a unique solution in $[-1,1]$. Let's call that unique solution $\omega_{Y, S}(A, B)$. Finally, I can write:
\[ \mathbb P(Y \in A, S \in B) = \Lambda_{2} \Bigg( \Lambda^{-1}(\mathbb P(Y \in A)), \Lambda^{-1}(\mathbb P(S \in B)), \omega_{Y,S}(A, B) \Bigg) \]
\begin{align*}
Cov(1_{\{Y\in A\}}, 1_{\{S \in B\}}) &= \mathbb{E}(1_{\{Y\in A\}},  1_{S\in B}   ) - \mathbb{E}(1_{\{Y\in A\}})\mathbb{E}(1_{\{S\in B\}}) \\
&= \mathbb P(Y \in , S \in B) - \mathbb P(Y \in A)P(S \in B) \\
&= \Lambda_{2} \Bigg( \Lambda^{-1}(\mathbb P(Y \in A)), \Lambda^{-1}(\mathbb P(S \in B)), \omega_{Y,S}(A, B) \Bigg) - \mathbb P(Y \in A)\mathbb P(S \in B) \\
&= G(\omega_{Y,S}(A, B)) - G(0)
\end{align*}
Since $G$ is strictly increasing, $\omega_{Y,S}(A, B)$ has the same sign as $Cov(1_{\{Y \in A\}}, 1_{\{S \in B\}})$

\subsection{Proof of Theorem 1}

Under mild interiority conditions, specifically, that all observed probabilities $ p_{k,z} \in (0,1) $ and that the solution satisfies $ |\omega_k| < 1 $,  the system of equations
$$
p_{k,z} = \Lambda_2\big(\lambda_k(\mu), \nu_z, \omega_k\big), \quad \text{for } k=1,\dots,q-1,\ z=0,1
$$
has exactly one solution for the unknown parameters: $ \mu = (\mu_1, \dots, \mu_{q-1}) \in \mathbb{R}^{q-1} $, with baseline $ \mu_q = 0 $, $ \omega = (\omega_1, \dots, \omega_{q-1}) \in (-1,1)^{q-1} $. I are given $ 2(q-1) $ equations (two per category$ k $: one for $ z=0 $, one for $ z=1 $). Each equation has the form:

$$
p_{k,z} = \frac{1}{1 + e^{-\lambda_k(\mu)} + e^{-\nu_z} + (1 - \omega_k) e^{-\lambda_k(\mu) - \nu_z}}
$$
Here:$ \nu_z = \log\left( \frac{\mathbb{P}(S=1 \mid Z=z)}{1 - \mathbb{P}(S=1 \mid Z=z)} \right) $ is known (given by data).  $ \lambda_k(\mu) = \log\left( \frac{\pi_k(\mu)}{1 - \pi_k(\mu)} \right) $, where $ \pi_k(\mu) = \frac{e^{\mu_k}}{\sum_{j=1}^q e^{\mu_j}} $, and $ \mu_q = 0 $. So $ \lambda_k(\mu) $ depends on $ \mu $, and $ \omega_k $ is a separate parameter per class. Our goal: Show that only one pair $ (\mu, \omega) $ satisfies all equations.\\

\textbf{Step 1: Solve for $ \lambda_k^{\ast} (\mu)$}

For each category $ k $, use both $ z=0 $ and $ z=1 $ to do it :
$$
p_{k,0} = \Lambda_2(\lambda_k(\mu), \nu_0, \omega_k), \quad p_{k,1} = \Lambda_2(\lambda_k(\mu), \nu_1, \omega_k)
$$
Crucially, both share the same $ \lambda_k(\mu) $ and $ \omega_k $. I can solve each equation for $ \omega_k $: from the formula:

$$
\Lambda_2(u,v,\omega) = \frac{1}{1 + e^{-u} + e^{-v} + (1-\omega)e^{-u-v}} = p
\Rightarrow
\omega = 1 - e^{u+v} \left( \frac{1}{p} - 1 - e^{-u} - e^{-v} \right)
$$
Apply this to both $ z=0 $ and $ z=1 $:

$$
\omega_k = 1 - e^{\lambda_k(\mu) + \nu_0} \left( \frac{1}{p_{k,0}} - 1 - e^{-\lambda_k(\mu)} - e^{-\nu_0} \right)
$$
$$
\omega_k = 1 - e^{\lambda_k(\mu) + \nu_1} \left( \frac{1}{p_{k,1}} - 1 - e^{-\lambda_k(\mu)} - e^{-\nu_1} \right)
$$

Set them equal:
$$
e^{\lambda_k(\mu) + \nu_0} \left( \frac{1}{p_{k,0}} - 1 - e^{-\lambda_k(\mu)} - e^{-\nu_0} \right)
=
e^{\lambda_k(\mu) + \nu_1} \left( \frac{1}{p_{k,1}} - 1 - e^{-\lambda_k(\mu)} - e^{-\nu_1} \right)
$$

Simplify both sides:

Left: $ e^{\lambda_k(\mu)} e^{\nu_0} \left( \frac{1}{p_{k,0}} - 1 \right) - e^{\nu_0} - e^{\lambda_k(\mu)} $

Right: $ e^{\lambda_k(\mu)} e^{\nu_1} \left( \frac{1}{p_{k,1}} - 1 \right) - e^{\nu_1} - e^{\lambda_k(\mu)} $

Subtract $ e^{\lambda_k(\mu)} $ from both sides and rearrange:

$$
e^{\lambda_k(\mu)} \left[ e^{\nu_0} \left( \frac{1}{p_{k,0}} - 1 \right) - e^{\nu_1} \left( \frac{1}{p_{k,1}} - 1 \right) \right] = e^{\nu_0} - e^{\nu_1}
$$

Therefore:

$$
e^{\lambda_k(\mu)} = \frac{ e^{\nu_0} - e^{\nu_1} }{ e^{\nu_0} \left( \frac{1}{p_{k,0}} - 1 \right) - e^{\nu_1} \left( \frac{1}{p_{k,1}} - 1 \right) }
$$

This is a constant determined entirely by the observed $ p_{k,0}, p_{k,1} $ and known $ \nu_0, \nu_1 $. So for each $ k $, $ \lambda_k(\mu) $ is uniquely determined. Call this known value $ \lambda_k^{\ast} $.\\

\textbf{Step 2: From $ \lambda_k^{\ast} $, recover $ \mu $}

Recall:
$$
\lambda_k(\mu) = \log\left( \frac{e^{\mu_k}}{ \sum_{j \ne k} e^{\mu_j} } \right)
$$

Since $ \mu_q = 0 $, the sum $ \sum_{j \ne k} e^{\mu_j} $ includes $ e^{\mu_q} = 1 $.

Let’s define:  $ x_k = e^{\mu_k} > 0 $ for $ k = 1, \dots, q-1 $  and  $ x_q = 1 $

Then:

$$
\lambda_k^* = \log\left( \frac{x_k}{ \sum_{j \ne k} x_j } \right)
\Rightarrow
e^{\lambda_k^*} = \frac{x_k}{ \sum_{j \ne k} x_j }
$$

Let $ s = x_1 + \dots + x_{q-1} $, so $ \sum_{j \ne k} x_j = s + 1 - x_k $

Then:

$$
e^{\lambda_k^*} = \frac{x_k}{s + 1 - x_k}
\Rightarrow
x_k = e^{\lambda_k^*} (s + 1 - x_k)
\Rightarrow
x_k (1 + e^{\lambda_k^*}) = e^{\lambda_k^*} (s + 1)
\Rightarrow
x_k = \frac{ e^{\lambda_k^*} (s + 1) }{ 1 + e^{\lambda_k^*} }
$$

Now sum over $ k = 1 $ to $ q-1 $:

$$
s = \sum_{k=1}^{q-1} x_k = (s + 1) \sum_{k=1}^{q-1} \frac{ e^{\lambda_k^*} }{ 1 + e^{\lambda_k^*} }
$$

Let $ \beta = \sum_{k=1}^{q-1} \frac{ e^{\lambda_k^*} }{ 1 + e^{\lambda_k^*} } $. This is a known number (since each $ \lambda_k^* $ is known).

So:

$$
s = (s + 1) \beta
\Rightarrow
s(1 - \beta) = \beta
\Rightarrow
s = \frac{\beta}{1 - \beta}
$$

This is well-defined and unique, as long as $ \beta < 1 $. $ \beta < 1 $ because: $ \frac{ e^{\lambda_k^*} }{ 1 + e^{\lambda_k^*} } = \mathbb{P}(Y = c_k) $ under the model, and since there’s also category$ q $, with probability $ \mathbb{P}(Y = c_q) = \frac{1}{s + 1} > 0 $, we have $ \beta = 1 - \mathbb{P}(Y = c_q) < 1 $. So $s$  is uniquely determined. Then, for each $k$:

$$
x_k = \frac{ e^{\lambda_k^*} (s + 1) }{ 1 + e^{\lambda_k^*} } \quad \text{(uniquely determined)}
\Rightarrow
\mu_k = \log x_k \quad \text{(uniquely determined)}
$$

So $ \mu $ is uniquely recovered from the $ \lambda_k^{\ast} $, which came from the data.

\textbf{Step 3: Recover $ \omega_k $}

Now that $ \mu $ is known, $ \lambda_k(\mu) $ is known (in fact, it’s $ \lambda_k^* $).

Plug into either equation (say, $ z=0 $):

$$
\omega_k = 1 - e^{\lambda_k(\mu) + \nu_0} \left( \frac{1}{p_{k,0}} - 1 - e^{-\lambda_k(\mu)} - e^{-\nu_0} \right)
$$

This gives a unique value for $ \omega_k $, and by construction, it will also satisfy the equation for $ z=1 $ (because I derived $ \lambda_k(\mu) $ by enforcing consistency between the two). Under interiority ($ p_{k,z} \in (0,1) $), this $ \omega_k $ will lie in $ (-1,1) $, as required.

we have shown that : i) For each category$ k $, the value $ \lambda_k(\mu) $ is uniquely pinned down by the two observations $ p_{k,0}, p_{k,1} $ and known $ \nu_0, \nu_1 $. ii)  From the full set $ \{\lambda_k(\mu)\}_{k=1}^{q-1} $, I uniquely recover $ \mu \in \mathbb{R}^{q-1} $ via normalization with $ \mu_q = 0 $. ii) Then, each $ \omega_k $ is uniquely recovered from the model equation.

Therefore, under the stated interiority conditions, there is exactly one solution $ (\mu, \omega) $ to the system.

\subsection{Proof of theorem 2}
\label{app:asym}

To derive the asymptotic distribution of the two-step estimator (\cite{newey1994large}) \(\hat{\theta} = (\hat{\theta}_1, \dots, \hat{\theta}_{q-1})\), where \(\theta_k = (\beta_k, \gamma_k)\), we follow a standard approach for two-step M-estimators under regularity conditions. The key idea is that estimation error from Step 1 (the selection equation) propagates into Step 2, and this must be accounted for in the asymptotic variance of \(\hat{\theta}\). Let the true parameters be \(\delta_0\) (from Step 1) and \(\theta_0 = (\theta_{10}, \dots, \theta_{(q-1)0})\) (from Step 2), where   \(\theta_{k0} = (\beta_{k0}, \gamma_{k0})\), \(\beta_{q0} = 0\) by normalization.

\noindent
\textbf{Step 1}: \(\hat{\delta}\) solves the standard logit score equation:
\[
\frac{1}{n} \sum_{i=1}^n \psi_1(W_i, S_i; \delta) = 0,
\quad \text{where} \quad 
\psi_1(W_i, S_i; \delta) = \big(S_i - \Lambda(W_i'\delta)\big) W_i.
\]
Under standard regularity conditions (correct specification, identification, smoothness, etc.), we have:
\[
\sqrt{n}(\hat{\delta} - \delta_0) \xrightarrow{d} N\big(0,\, A_1^{-1} \big),
\]
with
\[
A_1 = \mathbb{E}\big[ \Lambda(W'\delta_0)(1 - \Lambda(W'\delta_0)) W W' \big].
\]

\noindent
\textbf{Step 2}: Define for each observation the contribution to the second-step log-likelihood:
\[
\ell_i(\theta, \delta) = \sum_{k=1}^{q-1} S_i \cdot \mathbf{1}\{Y_i = c_k\} \cdot 
\log \Lambda_2\Big( \eta_{ik}(\beta), W_i'\delta, \omega_{ik}(\gamma_k) \Big),
\]
where: \(\eta_{ik}(\beta) = X_i'\beta_k - \log\big( \sum_{j \neq k} e^{X_i'\beta_j} \big)\),  \(\omega_{ik}(\gamma_k) = \tanh(X_i'\gamma_k)\).

Then the second-step estimator solves:
\[
\frac{1}{n} \sum_{i=1}^n \psi_2(Z_i; \theta, \hat{\delta}) = 0,
\quad \text{with} \quad 
\psi_2(Z_i; \theta, \delta) = -\nabla_\theta \ell_i(\theta, \delta),
\]
where \(Z_i = (Y_i, S_i, X_i, Z_i)\) (note \(W_i = (X_i, Z_i)\)).

Let the population second-step criterion be:
\[
Q_2(\theta, \delta) = \mathbb{E}\big[ \ell_i(\theta, \delta) \big].
\]
Assume this is uniquely maximized at \((\theta_0, \delta_0)\). Write the first-order condition for \(\hat{\theta}\):
\[
0 = \frac{1}{n} \sum_{i=1}^n \psi_2(Z_i; \hat{\theta}, \hat{\delta}).
\]
Add and subtract the population expectation and expand around \((\theta_0, \delta_0)\):
\begin{eqnarray*}
  &  0 = \underbrace{\mathbb{E}[\psi_2(Z; \theta_0, \delta_0)]}_{=0} 
+ \frac{1}{n} \sum_{i=1}^n \big( \psi_2(Z_i; \theta_0, \delta_0) - \mathbb{E}[\psi_2] \big)
+  \nabla_\theta \mathbb{E}[\psi_2(Z; \bar{\theta}, \bar{\delta})]
(\hat{\theta} - \theta_0)
+ \\ & \nabla_\delta \mathbb{E}[\psi_2(Z; \tilde{\theta}, \tilde{\delta})]
(\hat{\delta} - \delta_0)
+ o_p(n^{-1/2}),
\end{eqnarray*}

for some \(\bar{\theta}, \tilde{\theta}\) between \(\hat{\theta}\) and \(\theta_0\), and similarly for \(\delta\).
Define: \(H_2 = -\mathbb{E} \big[ \nabla_\theta \psi_2(Z; \theta_0, \delta_0) \big] = \mathbb{E} \big[ \nabla_{\theta\theta}^2 \ell_i(\theta_0, \delta_0) \big]\): Hessian (information matrix for Step 2), \(G = \mathbb{E} \big[ \nabla_\delta \psi_2(Z; \theta_0, \delta_0) \big] = -\mathbb{E} \big[ \nabla_{\theta\delta}^2 \ell_i(\theta_0, \delta_0) \big]\): cross-derivative capturing how Step 1 affects Step 2.
Then:
\[
\sqrt{n}(\hat{\theta} - \theta_0) = H_2^{-1} \left[ \frac{1}{\sqrt{n}} \sum_{i=1}^n \psi_2(Z_i; \theta_0, \delta_0) + G (\hat{\delta} - \delta_0) \sqrt{n} \cdot \frac{1}{\sqrt{n}} \right] + o_p(1)
\]
\[
= H_2^{-1} \left[ \frac{1}{\sqrt{n}} \sum_{i=1}^n \psi_2(Z_i; \theta_0, \delta_0) + G \cdot \sqrt{n}(\hat{\delta} - \delta_0) \right] + o_p(1).
\]
Now substitute the asymptotic for \(\sqrt{n}(\hat{\delta} - \delta_0)\):
\[
\sqrt{n}(\hat{\delta} - \delta_0) = \frac{1}{\sqrt{n}} \sum_{i=1}^n I_1^{-1} \psi_1(W_i, S_i; \delta_0) + o_p(1),
\]
since the logit MLE is asymptotically linear with influence function \(I_1^{-1} \psi_1\). Thus,
\[
\sqrt{n}(\hat{\theta} - \theta_0) = H_2^{-1} \left[ \frac{1}{\sqrt{n}} \sum_{i=1}^n \psi_2(Z_i; \theta_0, \delta_0) + G I_1^{-1} \cdot \frac{1}{\sqrt{n}} \sum_{i=1}^n \psi_1(W_i, S_i; \delta_0) \right] + o_p(1).
\]
Define the combined influence function:
\[
\phi(Z_i) = \psi_2(Z_i; \theta_0, \delta_0) + G I_1^{-1} \psi_1(W_i, S_i; \delta_0).
\]
Then:
\[
\sqrt{n}(\hat{\theta} - \theta_0) \xrightarrow{d} N\big(0,\, H_2^{-1} \Omega H_2^{-1}\big),
\]
where
\[
\Omega = \mathbb{E}\big[ \phi(Z) \phi(Z)' \big]
= \mathbb{E}\left[ \left( \psi_2 + G I_1^{-1} \psi_1 \right) \left( \psi_2 + G I_1^{-1} \psi_1 \right)' \right].
\]
If the model is correctly specified and the second-step likelihood is smooth, then under standard regularity conditions (interchange of differentiation and integration, identifiability, etc.), this simplifies.
Under standard regularity conditions (i.i.d. sampling, correct specification of both selection and outcome models, identification, smoothness, and non-singularity of information matrices), the two-step estimator \(\hat{\theta}\) satisfies:
\[
\sqrt{n}(\hat{\theta} - \theta_0) \xrightarrow{d} N\big(0,\, V_\theta \big),
\]
where the asymptotic variance is
\[
V_\theta = H_2^{-1} \left[ \mathbb{E}\big[ \psi_2 \psi_2' \big] + G I_1^{-1} G' \right] H_2^{-1},
\]
with:\(H_2 = -\mathbb{E}[ \nabla_{\theta\theta}^2 \ell_i(\theta_0, \delta_0) ]\), \(G = -\mathbb{E}[ \nabla_{\theta\delta}^2 \ell_i(\theta_0, \delta_0) ]\), \(I_1 = \mathbb{E}[ \psi_1 \psi_1' ] = \mathbb{E}[ \Lambda(W'\delta_0)(1 - \Lambda(W'\delta_0)) W W' ]\).

Interpretation: The first term \(\mathbb{E}[\psi_2 \psi_2']\) is the variance that would arise if \(\delta_0\) were known. The second term \(G I_1^{-1} G'\) is the variance inflation due to estimating \(\delta\) in the first step.  If the selection mechanism is independent of the outcome (i.e., \(G = 0\)), then the two-step estimator is asymptotically equivalent to the full MLE.

\section{Wealth index}
\label{asset}
We also have information on the sets of assets possessed by the household, and we use these variables to construct a simple asset index. This is done in a straightforward manner for physical assets, by a simple weighted average of the binary responses corresponding to whether or not a given list of items are owned by a household, using weights from a principal components analysis (PCA).
\[asset = \sum_{i=1}^{n}\omega_i b_i\]
We use 18 binary variables in this application: wall in good material, roof in good material, floor in good material, access to drinking water in the dry season and in the rainy season, access to electricity, use of the electricity network, clean restrooms, good waste management, good waste water management, owns a TV, has an iron, has a refrigerator, has a stove, has a computer, has cable, owns a car.

\label{app:}


	
\end{document}